\documentclass[preprint,showpacs,preprintnumbers,superscriptaddress,amsmath,amssymb]{revtex4}

\usepackage{graphicx}
\usepackage{dcolumn}
\usepackage{bm}
\usepackage{color}

\begin{document}
\title{Impurity clustering and impurity-induced bands in PbTe-, SnTe-, and GeTe-based bulk thermoelectrics}
\author{Khang Hoang}
\affiliation{Materials Department, University of California, Santa Barbara, California 93106, USA}
\author{S. D. Mahanti}%
\affiliation{Department of Physics and Astronomy, Michigan State University, East Lansing, Michigan 48824, USA}
\author{Mercouri G. Kanatzidis}
\affiliation{Department of Chemistry, Northwestern University, Evanston, Illinois 60208, USA}
\affiliation{Materials Science Division, Argonne National Laboratory, Argonne, Illinois 60439, USA}

\date{\today}

\begin{abstract}

Complex multicomponent systems based on PbTe, SnTe, and GeTe are of great interest for infrared devices and high-temperature thermoelectric applications. A deeper understanding of the atomic and electronic structure of these materials is crucial for explaining, predicting, and optimizing their properties, and to suggest new materials for better performance. In this work, we present our first-principles studies of the energy bands associated with various monovalent (Na, K, and Ag) and trivalent (Sb and Bi) impurities and impurity clusters in PbTe, SnTe, and GeTe using supercell models. We find that monovalent and trivalent impurity atoms tend to come close to one another and form impurity-rich clusters, and the electronic structure of the host materials is strongly perturbed by the impurities. There are impurity-induced bands associated with the trivalent impurities that split off from the conduction-band bottom with large shifts towards the valence-band top. This is due to the interaction between the $p$ states of the trivalent impurity cation and the divalent anion which tends to drive the systems towards metallicity. The introduction of monovalent impurities (in the presence of trivalent impurities) significantly reduces (in PbTe and GeTe) or slightly enhances (in SnTe) the effect of the trivalent impurities. One, therefore, can tailor the band gap and band structure near the band gap (hence transport properties) by choosing the type of impurity and its concentration or tuning the monovalent/trivalent ratio. Based on the calculated band structures, we are able to explain qualitatively the measured transport properties of the whole class of PbTe-, SnTe-, and GeTe-based bulk thermoelectrics.

\end{abstract}

\maketitle

\section{Introduction}

The thermoelectric phenomenon involves direct thermal-to-electric energy conversion, which can be used for both refrigeration and power generation. Together with other technologies for energy generation and conversion, it is expected to play an increasingly important role in meeting the energy demands of the next generations.~\cite{CMMP} Although the efficiency of thermoelectric materials have been significantly improved over the last decade,~\cite{sootsman,MRSthermo2} major advances are still needed to meet the future requirements. A fundamental understanding of these materials is crucial for explaining, predicting, and optimizing their properties, and also to suggest new materials for thermoelectric applications. First-principles calculations based on density functional theory (DFT) have proved to be extremely helpful in this regard.~\cite{MRSDFT}

The efficiency of a thermoelectric device is controlled by the dimensionless thermoelectric figure of merit, $ZT$, defined as
\begin{equation}\label{ZT}
ZT=\frac{\sigma S^{2}}{\kappa}T,
\end{equation}
where $\sigma$ is the electrical conductivity, $S$ is the thermopower (Seebeck coefficient), $T$ is the operating temperature, and $\kappa=\kappa_{el}+\kappa_{latt}$ is the total thermal conductivity containing an electronic part $\kappa_{el}$ (zero-current electronic thermal conductivity) and a lattice part $\kappa_{latt}$; $\sigma S^{2}$ is known as the thermoelectric ``power factor'' and depends primarily on the electronic structure. Within Boltzmann transport equation approach, the transport coefficients $\sigma$ and $S$ for a cubic system can be calculated using the following
equations:~\cite{mahan-sofo}
\begin{equation}\label{eq;elecconductivity}
  \sigma=e^{2}\int^{+\infty}_{-\infty}d\epsilon(-\frac{\partial f_{0}}{\partial \epsilon})\Sigma (\epsilon),
\end{equation}
\begin{equation}\label{eq;thermopower}
   S=\frac{e}{T\sigma}\int^{+\infty}_{-\infty}d\epsilon(-\frac{\partial f_{0}}{\partial \epsilon})\Sigma (\epsilon) (\epsilon -\mu),
\end{equation}
where $\mu$ is the chemical potential, $e$ the electron charge, $f_{0}$ is the Fermi-Dirac distribution function, and the transport distribution function $\Sigma(\epsilon)$ is given by
\begin{equation}\label{eq;transportfunction}
\Sigma (\epsilon)=\sum_{\mathbf{k}}\nu_{x}(\mathbf{k})^{2} \tau(\mathbf{k})\delta(\epsilon-\epsilon(\mathbf{k})).
\end{equation}
In Eq.~(\ref{eq;transportfunction}), the summation is over the first Brillouin zone (BZ), $\nu_{x}(\mathbf{k})$ is the group velocity of the carriers with wavevector $\mathbf{k}$ along the direction ($x$) of the applied field, $\tau(\mathbf{k})$ is the relaxation time, and $\epsilon(\mathbf{k})$ is the dispersion relation for the carriers; the band index is omitted for simplicity. When the relaxation time depends on $\mathbf{k}$ through $\epsilon(\mathbf{k})$, the transport distribution function takes the form~\cite{mahan-sofo}
\begin{equation}\label{eq;parabolic}
  \Sigma(\epsilon)=N(\epsilon)\nu_{x}(\epsilon)^{2}\tau(\epsilon),
\end{equation}
where $N(\epsilon)$ is the electronic density of states (DOS).

As $ZT$ approaches $\infty$, the thermoelectric conversion efficiency approaches its Carnot value.~\cite{mahan-sofo} Increasing $ZT$ has been one of the most challenging tasks (increasing $\sigma$ usually reduces $S$ and increases $\kappa_{el}$, the net result being a reduction in $ZT$), although there are no fundamental thermodynamic arguments against achieving very high values of $ZT$.~\cite{mahanSSP} More realistic approaches to increase $ZT$ have followed two different paths. One is to reduce $\sigma$ and $\kappa_{el}$ and increase $S$ by manipulating the DOS near the chemical potential through band-structure engineering or strong correlation effects. The other is to decrease $\kappa_{latt}$ through lattice engineering. Examples of the former are In-doped PbTe,~\cite{gelbstein} Tl-doped PbTe,~\cite{heremans} and doped cobaltates.~\cite{takahata} Those of the latter are skutterudites~\cite{nolasARMS} and AgPb$_{m}$SbTe$_{m+2}$ (LAST-$m$; LAST stands for Lead Antimony Silver Tellurium) containing Ag-Sb-rich nanoscale domains.~\cite{LAST} Recently, it has been suggested that one can also achieve high $ZT$ in systems such as In$_{4}$Se$_{3-\delta}$ by exploiting the intrinsic nanostructural bulk properties induced by charge-density waves.~\cite{rhyee}

Discoveries of new thermoelectric materials that can give large $ZT$ values, particularly at high temperatures ($T$$\sim$600$-$700 K), have created a great deal of excitement. LAST-$m$, which can be considered as a mixture of PbTe and AgSbTe$_{2}$, is among these materials. This $n$-type thermoelectric gives $ZT$=1.7 at 700 K for $m$=18.~\cite{LAST} Compared to PbTe, LAST-18 shows reduced thermal conductivity. The increase in $ZT$ in LAST-$m$ has been ascribed to the decrease in $\kappa_{latt}$ resulting from nanostructuring in the system where high-resolution transmission electron microscopy images indicate inhomogeneities in the microstructure of these materials, showing nanoscale domains of a Ag-Sb-rich phase embedded in a PbTe matrix.~\cite{LAST,eric,wu} Other bulk thermoelectrics discovered more recently also have high $ZT$ values and are nanostructured. Examples are $p$-type Ag(Pb$_{1-y}$Sn$_{y}$)$_{m}$SbTe$_{m+2}$ (LASTT-$m$),~\cite{LASTT} $p$-type Ag$_{1-x}$SnSb$_{1+x}$Te$_{3}$ (TAST-$m$),~\cite{TAST} $n$-type Pb$_{1-x}$Sn$_{x}$Te-PbS,~\cite{pbsntes} $n$-type Ag$_{1-x}$Pb$_{m}$$M$Te$_{m+2}$ ($M$=Sb, Bi),~\cite{LBST} $p$-type Na$_{1-z}$Pb$_{m}$Sb$_{y}$Te$_{m+2}$ (SALT-$m$),~\cite{SALT} and $n$-type K$_{1-x}$Pb$_{m+\delta}$Sb$_{1+\gamma}$Te$_{m+2}$ (PLAT-$m$).~\cite{PALT}

It is well known that transport and optical properties of semiconductors are dominated by the electronic states in the neighborhood of the band gap. From Eqs.~(1)$-$(5), clearly large values of $ZT$ require large values of $S$ and $\sigma$, both of which depend sensitively on the nature of the electronic states near the band gap. Thus it is essential to understand the underlying physics of the band gap formation and the nature of the electronic states in its neighborhood before being able to explain, predict, and optimize the properties of the systems. One, however, needs to know their atomic structures. Unfortunately, there is little, if any, information about the detailed atomic structures of the above mentioned thermoelectrics.

Our approach to understand the properties of LAST-$m$ and similar systems is based on a defect perspective. As a first-order approximation to the real system, LAST-$m$, for example, can be considered as PbTe doped with equal amounts of Ag and Sb, i.e., Ag and Sb being treated as defects (impurities) in PbTe. This approximation is expected to be good at low Ag, Sb concentrations (i.e., large $m$ values, which are of practical interest). One then looks at the electronic states induced by these impurities and their effect on the transport properties. The concentrations of Ag and Sb atoms ($x$$\sim$5\% for $m$=18) are small enough such that starting from an impurity picture is justified and physically meaningful. Yet they are large enough such that the effects of impurities and impurity-impurity interaction (either directly or indirectly through the host) on the electronic structure near the band gap are significant.

The studies of electronic states associated with impurities in PbTe have been so far mostly based on the single-particle DOS.~\cite{mahanti,bilcPRL,hazama,ahmadPRL,ahmadPRB,hase,hoangPRB,physicaB} Mahanti and Bilc~\cite{mahanti} reported rather limited results on the band structure of doped PbTe, only for PbTe simultaneously doped with Ag and Sb. A detailed analysis of the band structures showing the impurity-induced bands obtained in first-principles calculations is presently lacking. In fact, band structure can provide us with more information on the electronic states, especially on the impurity-induced or impurity-modified electronic states (i.e., the energy bands associated with the impurity in the system) and how they are formed. In addition, the relationship between the transport properties and electronic structure is more subtle and a detailed band structure is needed in order to have a better understanding of the transport properties of the system. Moreover, a general understanding of defect states in narrow band-gap semiconductors is also extremely helpful in searching for materials with desired properties.

In this paper, we present our extensive first-principles studies of the band structures of PbTe, SnTe, and GeTe in the presence of monovalent (Na, K, and Ag) and trivalent (Sb and Bi) substitutional impurities on the cation (Pb, Sn, or Ge) sites and discuss how the transport properties of these systems can be understood in terms of the calculated band structures, particularly those features which are sensitive to the impurities. We also discuss how the impurities interact with each other leading to impurity clustering and local relaxations near the impurity atoms. The arrangements of this paper is as follows: In Sec.~II, we present supercell models for defect calculations and the calculational details. Impurity clustering and local relaxations near the impurity atoms in various systems are discussed in Sec.~III. Impurity-induced bands associated with different monovalent and trivalent impurities in PbTe, SnTe, and GeTe are presented in Sec.~IV. Also in this section, we show how the changes in the band structures in going from one system to the other can explain the experimental transport data qualitatively. We conclude this paper with a summary in Sec.~V.

\section{Theoretical Modeling}

Among the three IV-VI binary tellurides, PbTe and SnTe are known to crystallize in a NaCl-type structure with face-centered cubic (fcc) unit cell. GeTe also has a NaCl-type structure, but with a slight (rhombohedral) distortion due to a phase transition at low temperature.~\cite{madelung2004} Since the distortion is small, GeTe is assumed to have the NaCl-type structure in the current studies.

To understand how each impurity perturbs the electronic structure of the host and how two impurities in a pair interfere with each other, first-principles calculations for single impurities and impurity pairs in $R$Te, where $R$=\{Pb, Sn, Ge\}, were carried out using supercell models. These calculations mainly made use of (2$\times$2$\times$2) cubic supercells, which contains 64 atoms/cell and requires lattice constant doubling in all the three directions with respect to that of the bulk materials. For a single impurity $M$, the supercell corresponds to the composition $MR_{m+1}$Te$_{m+2}$, $m$=30. The composition in the supercell for an impurity pair ($M$,$M'$) is $MM'R_{m}$Te$_{m+2}$.

The two impurity atoms in a ($M$,$M'$) pair are arranged with different distances and different geometries {\it vis-\`{a}-vis} the intervening Te atoms. They are arranged as the first, second, third, fourth, or fifth nearest neighbors (n.n.) of one another on the Pb sublattice with the impurity-impurity distance taking values $a/\sqrt{2}$, $a$, $a\sqrt{3/2}$, $a\sqrt{2}$, and $a\sqrt{3}$, respectively, where $a$ is the theoretical lattice constant of the bulk materials. The calculated values of the lattice constant are $a$=6.55 {\AA} (PbTe), 6.40 {\AA} (SnTe), and 6.02 {\AA} (GeTe). This supercell model has been used by Bilc {\it et al.}~\cite{bilcPRL} and by Hazama {\it et al.}~\cite{hazama} for ($M$,$M'$)=(Ag,Sb) in PbTe. Besides the (2$\times$2$\times$2) supercells, we also made use of (3$\times$3$\times$3) supercells which contains 216 atoms/cell ($m$=106).

Structural optimization, total energy and electronic structure calculations were performed within the DFT formalism, using the generalized-gradient approximation (GGA)~\cite{GGA} and the projector-augmented wave method~\cite{PAW1,PAW2} as implemented in the VASP code.~\cite{VASP1,VASP2,VASP3} Scalar relativistic effects (mass-velocity and Darwin terms) and spin-orbit interaction (SOI) were included, except in ionic optimization. In this case, only the scalar relativistic effects were taken into account since we found that the inclusion of SOI did not have a significant influence on the atomic structure. For the (2$\times$2$\times$2) supercells, we used a $5\times5\times5$ Monkhorst-Pack~\cite{monkhorst-pack} $\mathbf{k}$-point mesh in the self-consistent run; whereas for the (3$\times$3$\times$3) ones, 5 $\mathbf{k}$-points were used in the irreducible BZ. The energy cutoff was set to 300 eV and convergence with respect to self-consistent iterations was assumed when the total energy difference between consecutive cycles was less than $10^{-4}$ eV.

It is known that there are fundamental issues with DFT and the use of supercell method in studying defects in semiconductors.~\cite{supercellmethod} These include the tendency of DFT-GGA to underestimate the band gaps of semiconductors (the so-called ``band-gap problem'') and the finite size and artificial periodicity of the supercells. The former problem is severe, e.g., in PbTe where the calculated band gap is smaller than the experimental one by more than 50\% the defect states may artificially overlap with the valence- and/or conduction-band edges.~\cite{IBgroupIII} Although supercell calculations using methods which go beyond DFT-GGA, such as screened-exchange local-density approximation,~\cite{sxLDA} hybrid functional approximation,~\cite{HSE} and $GW$ approximation,~\cite{aulber} may help fix this problem, they are, however, still not affordable computationally due to the large supercell sizes and the heavy elements present in our systems.~\cite{note}

\section{Impurity Clustering in Bulk Thermoelectrics}

As a first step towards understanding the energetics of nanostructuring in telluride-based bulk thermoelectrics, we investigated various impurity pairs in different pair configurations embedded in the host materials. The formation energy of an impurity pair as a function of the pair distance can help identify the most stable configuration energetically and provide valuable information on how the impurity atoms are likely to arrange themselves in the host under certain synthesis conditions.

The formation energy $E^{f}$ of a defect X in neutral charge state is defined as~\cite{vdw}
\begin{equation}\label{eq;Ef}
  E^{f}=E_{\mathrm{tot}}(\mathrm{X})-E_{\mathrm{tot}}(\mathrm{bulk})-\sum_{i}n_{i}\mu_{i},
\end{equation}
where $E_{\mathrm{tot}}(\mathrm{X})$ and $E_{\mathrm{tot}}(\mathrm{bulk})$ are the total energies of a supercell containing X and of a supercell containing only bulk materials; $\mu_{i}$ is the chemical potential of species $i$ (host atoms or impurity atoms) which corresponds to the energy of the reservoir with which atoms of species $i$ are being exchanged, and $n_{i}$ denotes the number of atoms of species $i$ that have been added ($n_{i}$$>$0) or removed ($n_{i}$$<$0) to create the defect. Since we are interested mostly in the relative formation energies of a given pair for different spatial configurations, the precise values of the chemical potentials are not important. For simplicity we have fixed their values to the total energy (per atom) of the bulk in their standard metallic states.

\subsection{Impurity Clustering}

\begin{figure*}
\hspace*{0.0in}
\includegraphics[width=4.0in,angle=0]{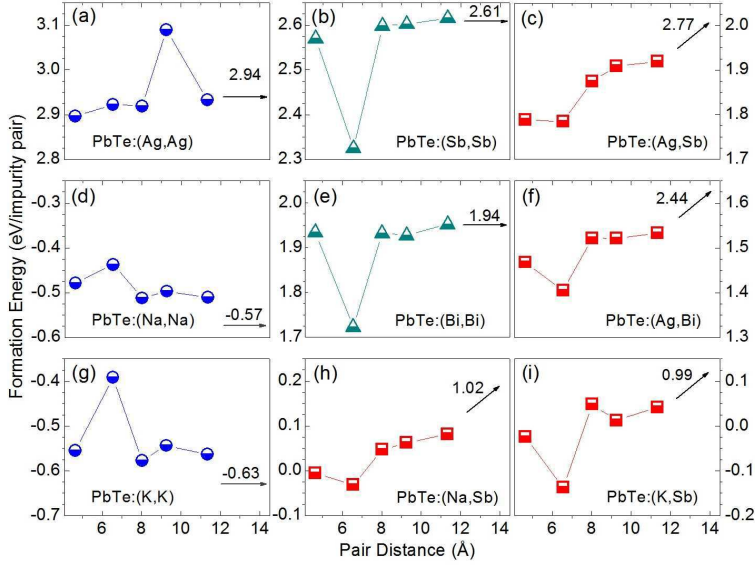} 
\vspace{-0.15in}
\caption{(Color online) Formation energies of various impurity pairs in PbTe as a function of the pair distance. The results were obtained in calculations using (2$\times$2$\times$2) supercells; SOI was not included (see the text). The values given with the arrows are the formation energies of the pairs at infinite pair distance (obtained by adding up the formation energies of the isolated impurities).}
\label{fig;EnePbTe}
\end{figure*}

Figures~\ref{fig;EnePbTe}(a)$-$\ref{fig;EnePbTe}(i) show the formation energy plots of X=(Ag,Ag), (Sb,Sb), (Ag,Sb), (Bi,Bi), (Ag,Bi), (Na,Na), (Na,Sb), (K,K), and (K,Sb) in PbTe as a function of the pair distance. The formation energies of the pairs at infinite pair distance are also given. These were obtained by adding up the formation energies of the isolated impurities. The pair binding energy ($E_{b}$) is calculated as the difference between the formation energy at a given pair distance and that at infinite pair distance.

We find that different impurity atoms behave quite differently. Let us look at the case when both the impurities are same (homo pairs). The monovalent and trivalent pairs show qualitatively different behavior. The monovalent alkali impurities tend to repel weakly whereas two monovalent Ag atoms and the trivalent pairs (Sb,Sb) and (Bi,Bi) tend to attract. There is a strong repulsion between two alkali atoms when they flank a Te atom. On the other hand, two Ag atoms repel most when they are the fourth n.n.~of each other.

The energy landscapes of (Sb,Sb) and (Bi,Bi) pairs are similar; see Figs.~\ref{fig;EnePbTe}(b) and \ref{fig;EnePbTe}(e). Both Sb and Bi, however, show completely different behavior compared to the monovalent impurities. The most notable feature is a large drop in $E^{f}$ at the second n.n.~distance. This feature is also apparent for pairs made from one monovalent atom and one trivalent atom (hetero pairs), they all have a minimum in $E^{f}$ at the second n.n.~distance; see Figs.~\ref{fig;EnePbTe}(c), \ref{fig;EnePbTe}(f), \ref{fig;EnePbTe}(h), and \ref{fig;EnePbTe}(i). For (Ag,Sb) pair, $E^{f}$ for the first and the second n.n.~distances are comparable, which is in agreement with the results reported by Hazama {\it et al.}~\cite{hazama} For all the hetero pairs we find the binding energy $E_{b}\sim$1.0 eV at the second n.n.~distance.

\begin{figure*}
\hspace*{0.0in}
\includegraphics[width=4.0in,angle=0]{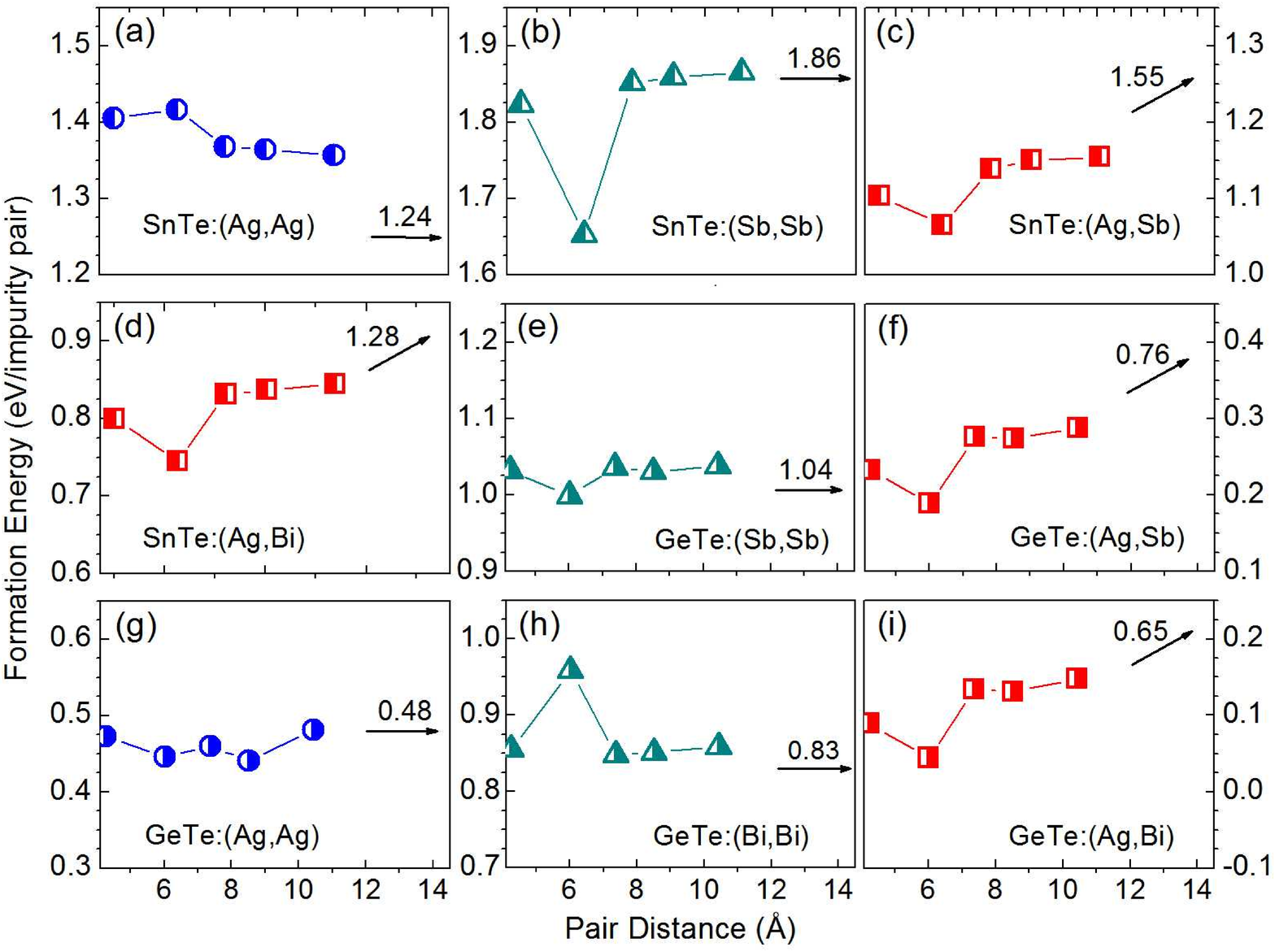} 
\vspace{-0.15in}
\caption{(Color online) Formation energy of various impurity pairs in SnTe and GeTe as a function of pair distance. The results were obtained in calculations using (2$\times$2$\times$2) supercells; SOI was not included (see the text). The values given with the arrows are the formation energies of the pairs at infinite pair distance (obtained by adding up the formation energies of the isolated impurities).}
\label{fig;EneSnTeGeTe}
\end{figure*}

For SnTe and GeTe, we have studied the impurity pairs made of monovalent Ag and/or trivalent Sb and Bi; see Figs.~\ref{fig;EneSnTeGeTe}(a)$-$\ref{fig;EneSnTeGeTe}(i). The formation energies of (Ag,Ag) in SnTe and GeTe do not change much as one varies the pair distance, although the Ag atoms in SnTe tend to repel each other or show a shallow minimum in GeTe. The (Sb,Sb) and (Bi,Bi) pairs in SnTe, on the other hand, have a large drop in $E^{f}$ at the second n.n.~distance, similar to what has been observed in PbTe. The minimum of $E^{f}$ at the second n.n.~distance for (Sb,Sb) in GeTe is much less pronounced. The formation energy of the (Bi,Bi) pair in GeTe, on the other hand, has a maximum at the second n.n.~distance.

Although the energy landscape of different homo pairs in SnTe and GeTe can be very different, any combination of one monovalent and one trivalent impurity has the lowest $E^{f}$ at the second n.n.~distance. This seems to be a robust characteristic of simultaneous doping of two impurities which are valence-compensated. The binding energy $E_{b}$ of the hetero pairs at the second n.n.~distance in SnTe and GeTe is $\sim$0.5 eV, which is smaller than in PbTe. Quantitative differences in the energy landscape of different impurity pairs in different host materials may result in different nanostructuring patterns. Of course, synthesis conditions should also play an important role in the formation of the embedded nano-domains.

We note that the above results were obtained in calculations where SOI was not included. Although the inclusion of SOI is not likely to change the energy landscapes we have presented, it may change the magnitude of the formation energy in some cases. Besides, one should be aware that SOI makes the band gap problem in PbTe and SnTe more serious by significantly reducing the gap and possibly causing an artificial overlap between the impurity-induced band and the valence band (see Sec.~IV). As a result, the total energies obtained from the calculations may not be accurate.

In order to see how the energy landscape looks like in larger supercell sizes, we carried out calculations for (Ag,Sb) and (Sb,Sb) pairs in PbTe using (3$\times$3$\times$3) supercells. Like in the case of (2$\times$2$\times$2) supercells, we also find that both (Ag,Sb) and (Sb,Sb) have a minimum in $E^{f}$ at the second n.n.~distance. For the (Ag,Sb) pair, $E^{f}$ at the second n.n.~distance is lower than that at the first, third, and fifth n.n.~distances by 41, 95, and 91 meV/pair; these numbers are 40, 90, and 135 meV/pair in (2$\times$2$\times$2) supercells. For (Sb,Sb), $E^{f}$ at the second n.n.~distance is lower than that at the first, third, and fifth n.n.~distances by 215, 222, and 232 meV/pair; these numbers are 245, 274, 291 meV/pair in (2$\times$2$\times$2) supercells. This suggests that, as far as the energy landscape is concerned, (2$\times$2$\times$2) supercells give reliable results.

To summarize, formation energy calculations of impurity pairs show that many impurity atoms in PbTe, SnTe, and GeTe tend to come close to one another and form impurity-rich clusters. This observation is consistent with our previous study of the systems using an ionic model,\cite{hoangMC} and with experiments where nanostructuring has been found in PbTe- and SnTe-based bulk thermoelectrics such as LAST-$m$,~\cite{LAST,eric,wu} Ag$_{1-x}$Pb$_{m}$$M$Te$_{m+2}$ ($M$=Sb and Bi),~\cite{LBST} TAST-$m$,~\cite{TAST} SALT-$m$,~\cite{SALT} and PLAT-$m$.~\cite{PALT} For GeTe-based systems such as (AgSbTe$_{2}$)$_{1-x}$(GeTe)$_{x}$, in addition to the solid-solution-like distribution of impurities~\cite{wood,plachkova} and microstructures (which were ascribed to twinning),~\cite{cook} {\it in situ} formed inhomogeneities and nanoscale domains were also reported.~\cite{yang} Our results are also in agreement with very recent first-principles studies for (Ag,Sb)-doped PbTe by Ke {\it et al.}~\cite{ke} that show: (i) Ag and Sb prefer to form Ag-Te-Sb-Te units along the (010) direction of the PbTe matrix and (ii) these units tend to form a maximal number of Ag-Sb pairings.

\subsection{Local Relaxations Caused by Impurity Clustering}

\begin{table}
\caption{Different bond lengths (in {\AA}) observed in PbTe simultaneously doped with monovalent and trivalent atoms. The two atoms in a pair are either the first, second, or third n.n.~of one another. Cases where bond lengths are different by $\sim$0.2 {\AA} or more in a given configuration are listed. The results were obtained in calculations using (2$\times$2$\times$2) supercells. For reference, Pb-Te bond length in bulk PbTe is 3.275 {\AA}.}
\label{tab;offcenter}
\begin{center}
\begin{tabular}{ccccc}
\hline \hline
 &&1st n.n.&2nd n.n.&3rd n.n.\\ \hline
(Ag,Sb)&Ag-Te&3.07, 3.20, 3.31&3.12, 3.41&- \\
&Sb-Te&2.93, 3.19, 3.44&-&2.96, 3.18, 3.38 \\ 
(Ag,Bi)&Ag-Te&-&3.14, 3.33&- \\
&Bi-Te&3.09, 3.22, 3.33&-&- \\ 
(Na,Sb)&Na-Te&-&3.26, 3.45&- \\
&Sb-Te&2.93, 3.18, 3.42&-&2.96, 3.18, 3.37 \\ 
(K,Sb)&K-Te&-&-&- \\
&Sb-Te&2.94, 3.18, 3.41&-&- \\
\hline \hline
\end{tabular}
\end{center}
\end{table}

Local geometry in the neighborhood of an impurity pair can be strongly distorted from an average structure (as measured in a diffraction measurement) when the two atoms in a pair are made of one monovalent and one trivalent atom, especially when they are close to one another. Local relaxation effects are relatively small when the impurities are far away from each other. We observe that some Te atoms that are the neighbors of the ($M$,$M'$) pair in PbTe and, in some cases, the impurity atoms themselves go ``off-center'' (i.e., not on the regular lattice sites). This results in two or more $M$-Te and $M'$-Te bond lengths. The off-centering occurs when the two atoms in a pair are the first, second, and/or third n.n.~of one another. In Table~\ref{tab;offcenter}, we list cases where bond lengths in a given pair configuration are different by $\gtrsim$0.2 {\AA}.

\begin{figure}
\includegraphics[width=2.5in,angle=0]{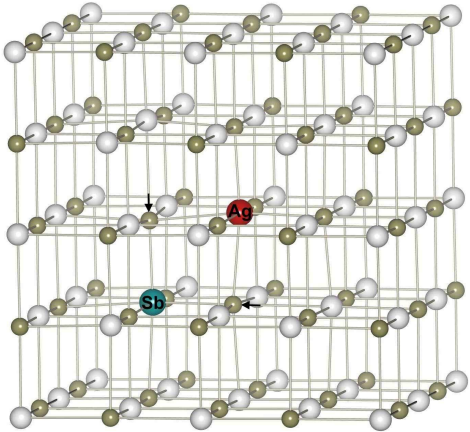} %
\caption{(Color online) Relaxed structure of (Ag,Sb)-doped PbTe where the two impurity atoms are the first nearest neighbors of one another in a (2$\times$2$\times$2) supercell. Large spheres are Ag (red) and Sb (blue), medium (gray) spheres Pb, and small (dark gray) spheres Te. The two Te atoms (marked by the arrows) in between Ag and Sb are shifted towards Sb by $\sim$0.2 {\AA}, resulting in different Ag-Te and Sb-Te bond lengths; see Table~\ref{tab;offcenter}.} \label{fig;Struct2x}
\end{figure}

\begin{figure}
\includegraphics[width=3.3in,angle=0]{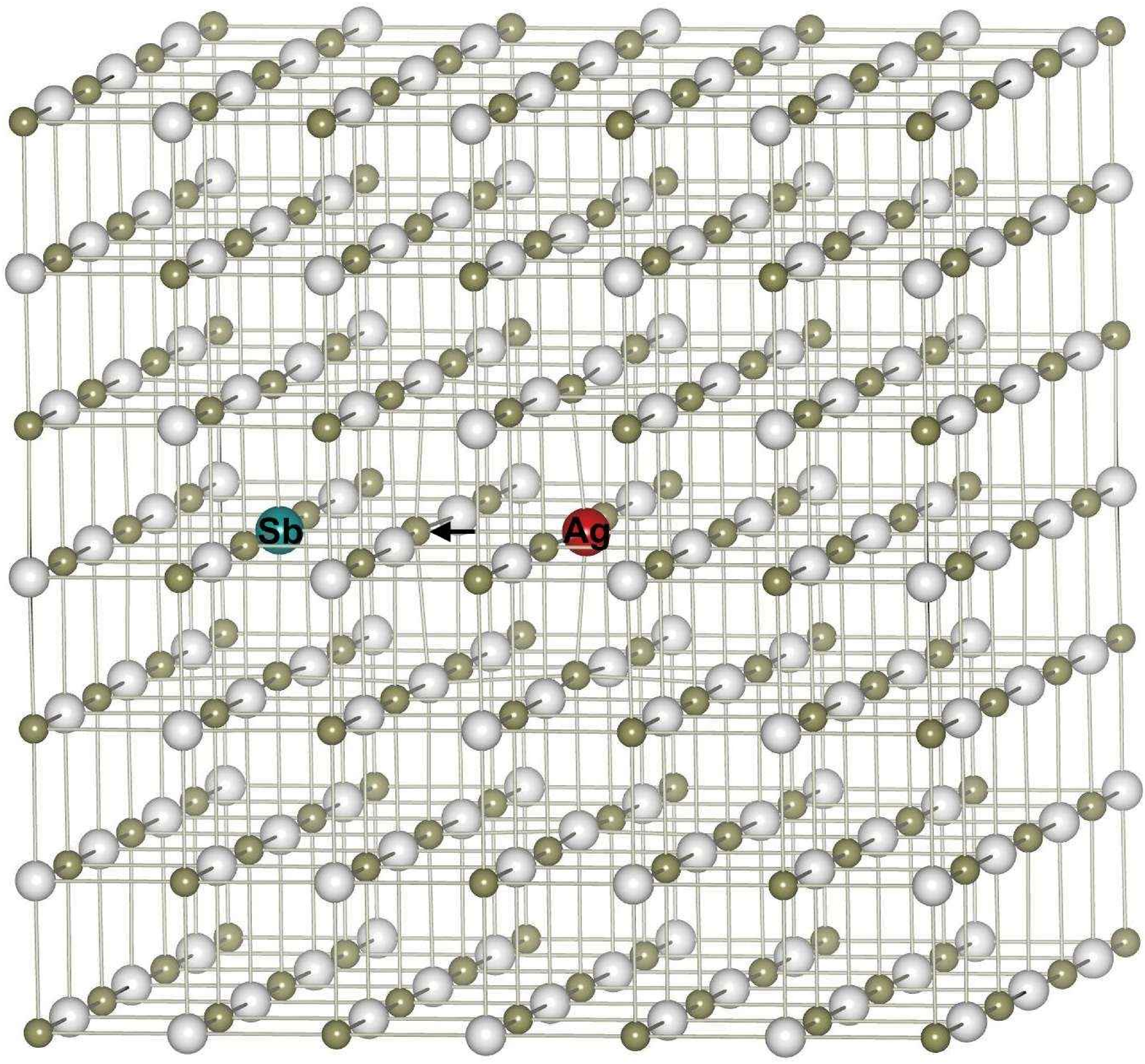} %
\caption{(Color online) Relaxed structure of (Ag,Sb)-doped PbTe where the two impurity atoms are the second nearest neighbors of one another in a (3$\times$3$\times$3) supercell. Large spheres are Ag (red) and Sb (blue), medium (gray) spheres Pb, and small (dark gray) spheres Te. The Te atom (marked by the arrow) in the Ag-Te-Sb chain is shifted towards Sb by $\sim$0.3 {\AA}, resulting in different Ag-Te and Sb-Te bond lengths.} \label{fig;Struct3x}
\end{figure}

Figure~\ref{fig;Struct2x} shows the relaxed structure of (Ag,Sb)-doped PbTe where the two impurity atoms are at the first n.n.~distance in a (2$\times$2$\times$2) supercell. We find that the two Te atoms in between Ag and Sb are shifted towards Sb by $\sim 0.2$ {\AA}, resulting in three Ag-Te bond lengths and three Sb-Te bond lengths (see Table~\ref{tab;offcenter}).

In the configuration where the two impurity atoms are at the second n.n.~distance in a (2$\times$2$\times$2) supercell, the Te atoms in the Ag-Te-Sb chain are slightly shifted toward Sb, resulting in two Ag-Te bond lengths (3.12 and 3.41 {\AA}). Note that in this case, Ag and Sb form an infinite Ag-Te-Sb-Te chain. This imposes an artificial constraint on the relaxation of Ag, Sb, and their neighboring Te atoms, and the bond lengths may not be given accurately. Our calculations using larger supercells indeed show that there are stronger relaxations. In Fig.~\ref{fig;Struct3x} we show the relaxed structure of (Ag,Sb)-doped PbTe where the two impurity atoms are at the second n.n.~distance in a (3$\times$3$\times$3) supercell. The Te atom in between Ag and Sb is shifted towards Sb by $\sim$0.3 {\AA}, resulting in three Ag-Te bond lengths (3.00, 3.10, and 3.74 {\AA}) and three Sb-Te bond lengths (2.98, 3.15, and 3.33 {\AA}).

Off-centering is also found with other impurity pairs and pair configurations in PbTe, SnTe, and GeTe. However, the bond length differences are usually smaller ($<$0.2 {\AA}). These small distortions may not be detected, for instance, in x-ray absorption fine structure (XAFS) analysis.~\cite{XAFS} The changes in the local bond length results from a combination of (i) the difference in the atomic radii of the impurity and the host (Pb, Sn, or Ge) atoms which causes the relaxation of the neighboring Te atoms and (ii) the strong and directional interaction between the trivalent impurity $p$ states and Te $p$ states (see Sec.~IV) which tends to pull Te atoms towards the trivalent impurity. These local distortions in the lattice geometry can potentially assist the formation of impurity-rich domains (e.g., Ag-Sb-rich nanoscale domains) in PbTe-, SnTe-, and GeTe-based systems. Besides, the off-centering observed in the systems is expected to have effects on their transport properties. The degree of off-centering may depend sensitively on the lattice constant (pressure).

Careful experimental studies are needed to confirm if there is off-centering and further experimental and theoretical studies are needed to understand the effects of off-centering on the transport properties of these systems.

\section{Impurity-Induced Bands}

The electronic structure of a semiconductor can be strongly disturbed in the presence of impurities. In this section, we present our comprehensive first-principles studies of the band structures of PbTe, SnTe, and GeTe doped with monovalent (Ag, Na, and K) and trivalent (Sb and Bi) impurities and discuss how different impurity-related properties of these systems can be understood in terms of the calculated band structures. The investigations focus mainly on the highest valence band and/or the lowest conduction band induced by the impurities, hereafter called {\it impurity-induced bands}, since they are most relevant to understanding the transport properties of these PbTe-, SnTe-, and GeTe-based bulk thermoelectrics. In the limit of extreme dilution, these bands approach the host bands.

\subsection{PbTe}

Before presenting the band structures of LAST-$m$ (PbTe doped with Ag and Sb) and other PbTe-based systems, let us summarize some of the important features of the band structure of undoped PbTe focusing on the highest valence band and the lowest conduction band. In PbTe, the valence $p$ states of Pb and Te play a dominant role in the formation of the valence and conduction bands. These bands are predominantly bonding and antibonding states of Te $p$ and Pb $p$ states. The conduction- and valence-band edges are almost symmetric through the band gap and both the maximum and the minimum occur at the same point in the $\mathbf{k}$ space.

In the fcc BZ, the direct band gap is at the $L$ point and the valence-band maximum (VBM) and the conduction-band minimum (CBM) are nondegenerate (disregarding spin degeneracy). The four inequivalent $L$ points in the fcc BZ are mapped into the $\Gamma$ point in the simple cubic (sc) BZ of the (2$\times$2$\times$2) supercell. The band extrema are, therefore, fourfold degenerate at $\Gamma$; see Fig.~\ref{fig;BandsPbTe}(a). With spin, the CBM (VBM) at $\Gamma$ has eightfold degeneracy. The band gap of PbTe gets reduced significantly (from 0.816 eV to 0.105 eV) in the presence of SOI due to the large lowering in energy of the Pb $p$ bands (dominant near the conduction-band bottom) and a smaller change in the Te $p$ bands (dominant near the valence-band top).~\cite{IBgroupIII} In the following sections, we will discuss how the eightfold degeneracy of the CBM and VBM is lifted in the presence of an impurity and what are its implications on the transport properties.

\subsubsection{Band structures of PbTe doped with Ag, Sb, and Bi}

\begin{figure}
\includegraphics[width=3.3in,angle=0]{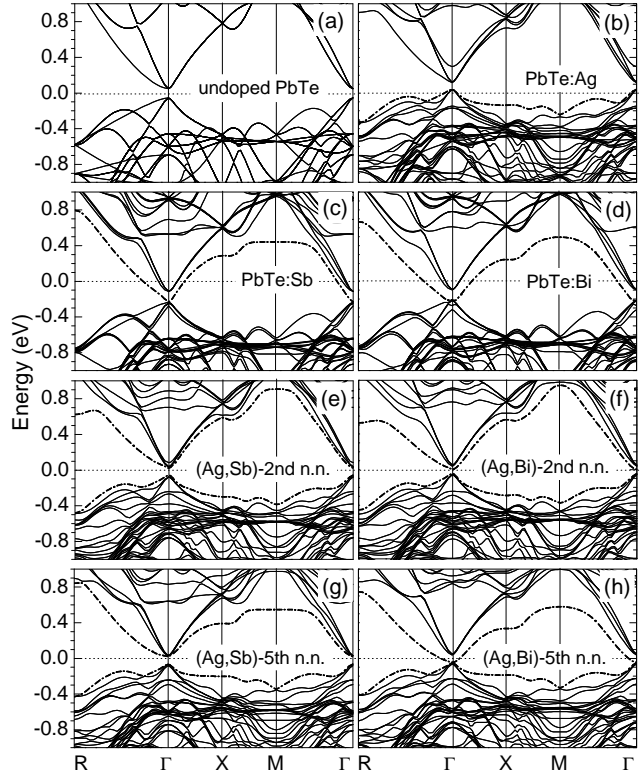} 
\vspace{-0.1in} \caption{Band structures of undoped PbTe and PbTe doped with Ag, Sb, Bi, (Ag,Sb), and (Ag,Bi). The impurity-induced bands are represented by the dash-dotted curves. The results were obtained in calculations using the (2$\times$2$\times$2) cubic supercell with SOI included. The two impurity atoms in a pair are either the second (2nd n.n.) or fifth (5th n.n.) nearest neighbors of one another. The Fermi level (0 eV) is set to the highest occupied states or in the band gap.} \label{fig;BandsPbTe}
\end{figure}

\begin{figure}
\includegraphics[width=3.2in,angle=0]{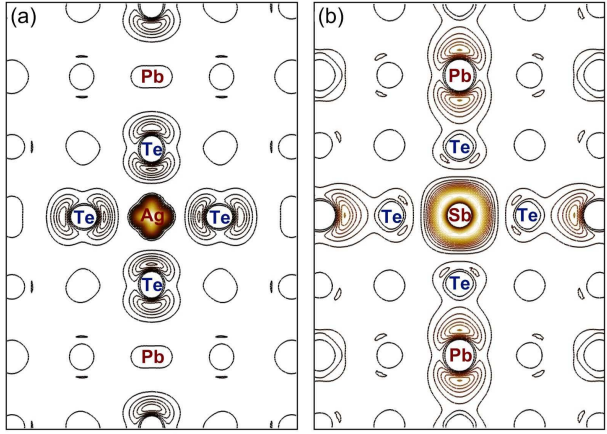} 
\caption{(Color online) Band-decomposed charge densities associated with the impurity-induced bands in PbTe doped with (a) Ag and (b) Sb. These partial densities were calculated from a preconverged wavefunction file for the specified bands. Contributions from all 10 $\mathbf{k}$-points in the irreducible BZ were included. The Ag-induced band is predominantly Ag $d$ states hybridizing with Te $p$ states, whereas the Sb-induced one is Sb $p$ and Pb $p$ states with some contribution from Te $p$ states.} \label{fig;IBandsAgnSb}
\end{figure}

First, we consider the effects of each impurity on the band structure separately. Figures~\ref{fig;BandsPbTe}(b)$-$\ref{fig;BandsPbTe}(d) show the band structures of $M$Pb$_{31}$Te$_{32}$ for $M$=Ag, Sb, or Bi. The highest Ag-induced band [the dash-dotted curve in Fig.~\ref{fig;BandsPbTe}(b)] is the nearly flat band (along $\Gamma$-$X$-$M$-$\Gamma$) splitting off from the rest of the valence band. This impurity-induced band overlaps with the states near the valence-band top. An examination of the partial charge density associated with this band shows that it is predominantly Ag $d$ states hybridizing with Te $p$ states, see Fig.~\ref{fig;IBandsAgnSb}(a). This band, therefore, can be identified with the resonant state in the single-particle electronic DOS which has been discussed earlier.~\cite{ahmadPRB} The Fermi level lies below the VBM (by $\sim$40 meV) indicating that the system is hole-doped. Note that the Ag $s$ state is high up in the conduction band. The Ag impurity at this concentration ($\sim$3\%) reduces the PbTe band gap from 105 meV to 73 meV. The conduction-band degeneracy (fourfold, without spin) is lifted near the CBM (at $\Gamma$) in the presence of the Ag impurity. However the lowest conduction band(s) are almost identical to that seen in pure PbTe.

The impurity-induced bands associated with Sb and Bi [the dash-dotted curves in Figs.~\ref{fig;BandsPbTe}(c) and \ref{fig;BandsPbTe}(d)] arise from strong interaction between the impurity $p$ level (which lies $\sim$0.6 eV above the CBM) and the conduction-band states. This results in the splitting of the fourfold degenerate conduction band of PbTe [the lowest conduction band in Figs.~\ref{fig;BandsPbTe}(c) and \ref{fig;BandsPbTe}(d)]. The CBM of PbTe, which originally has eightfold degeneracy (including spin) at $\Gamma$, now splits into a sixfold degenerate level (at $-$0.11 eV) and the twofold degenerate level (at $-$0.23 eV) which belongs to the impurity-induced band. Partial charge density analysis shows that the impurity-induced band associated with Sb is predominantly Sb $p$ hybridizing with Pb $p$ and some contribution from Te $p$ states; see Fig.~\ref{fig;IBandsAgnSb}(b). Similar characteristics are seen in the case of Bi.

For both Sb and Bi, the impurity-induced bands come down and close the band gap. Since they have one more electron than Pb, the Fermi level lies in the conduction band and is above the CBM by $\sim$114 meV (Sb) or by $\sim$94 meV (Bi), which indicates that the systems are electron-doped. Although the Sb(Bi)-induced band is a result of the Sb(Bi) $p$ and Pb $p$ interaction, it is the highly directional interaction between the Sb(Bi) $p$ states and the Te $p$ states that drives the system towards metallicity, a phenomenon that has been observed in many chalcogenides.~\cite{hoangPRL,Tlternaries,hoangthesis,pbs} The splitting between the impurity-induced band and the rest of the bands above it is affected by SOI and is, therefore, larger for Bi. This makes the Bi-induced band overlap with the VBM (of PbTe) near $\Gamma$, see Fig.~\ref{fig;BandsPbTe}(d). Both Sb and Bi leave the valence-band top almost unaffected except near the $\Gamma$ point.

As we mentioned in Sec.~II, the band gap of PbTe is underestimated by $\sim$50\% in DFT-GGA calculations. This means that the overlap between the Bi-induced band and the valence band as seen in Fig.~\ref{fig;BandsPbTe}(c) can be an artifact of DFT-GGA and that between the Sb-induced band and the valence band [Fig.~\ref{fig;BandsPbTe}(d)] there may be a finite energy gap. In addition, the curvature of the energy bands near the VBM and CBM might not be given correctly. In spite of these limitations, our first-principles calculations do show clear trends in the change of the electronic structure of PbTe when doped with different impurities whose concentration is of the order of a few percent. These findings can still be extremely useful in understanding the physical properties of these materials (as we will discuss below and in the next sections). Besides, it is not unreasonable to assume that the conduction band can be shifted rigidly when one makes corrections to the band gap. In that case, since Sb- and Bi-induced bands are derived from the lowest PbTe conduction-band states, they are expected to shift in the same direction with the conduction band, whereas the Ag-induced band, since it is split off from the valence-band top, is expected to go with the valence band.

\begin{figure}
\hspace*{-0.15in}
\includegraphics[width=3.0in,angle=270]{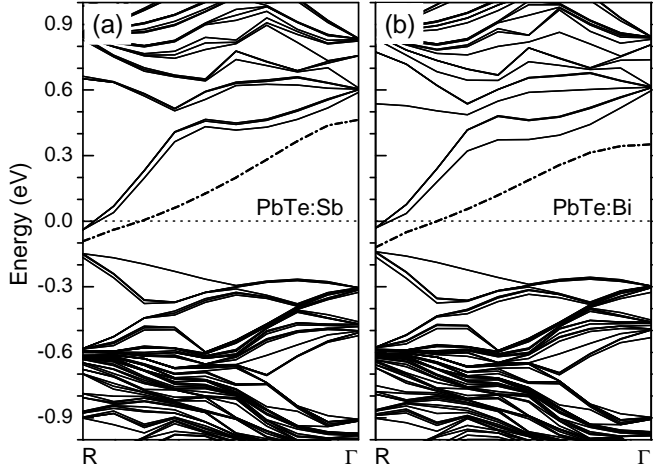} 
\vspace{-0.6in} \caption{Band structures of PbTe doped with (a) Sb and (b) Bi obtained in calculations using (3$\times$3$\times$3) supercells with SOI included. The impurity-induced bands are represented by the dash-dotted curves. Note that the zero of the energy may not be given accurately because the calculations scanned only a very small part of the BZ.} \label{fig;Bands3xSbnBi}
\end{figure}

To illustrate how the band structures change in going to lower impurity concentrations, we show in Figs.~\ref{fig;Bands3xSbnBi}(a) and \ref{fig;Bands3xSbnBi}(b) the band structures of PbTe doped with Sb and Bi obtained in calculations using (3$\times$3$\times$3) supercells (the impurity concentration being $\sim$1\%). For this supercell, the band extrema occur at the $R$ point of the cubic BZ. We find that there are finite band gaps between the impurity-induced bands and the VBM; 60 meV (in the case of Sb) and 20 meV (Bi). Splittings near the conduction-band bottom are smaller compared to the (2$\times$2$\times$2) supercell (the impurity concentration being $\sim$3\%), because the effect of the impurity on the host PbTe becomes weaker in the larger supercell. For example, the splitting between the Sb-induced band and the rest of the conduction band is 118 meV in the (2$\times$2$\times$2) supercell, but only 52 meV in the (3$\times$3$\times$3) one; for Bi, the splitting goes, respectively, from 178 meV to 88 meV. Changing the impurity concentration, therefore, does not only change the band gap but also alters the arrangement of the energy bands near the band gap region.

Our first-principles studies thus gives a physical picture of how different impurities affect the electronic structure of the host material. Clearly, a simple rigid band picture where one simply populates the host PbTe bands with electrons or holes is not appropriate for these PbTe-based systems, particularly for impurity concentrations $>$1 at\%. From a materials design perspective, this opens up an opportunity to tune the band gap and band structure in the neighborhood of the band gap (hence transport properties) by choosing the type of impurity and its concentration. One can also combine different types of impurities to achieve the desired properties.

Note that, although the trivalent impurities tend to come close and form some sort of impurity-rich cluster (see our discussions in Sec.~III), the average distance between impurities at the impurity concentration $\sim$1 at\% is relatively large and hence the direct impurity-impurity interaction is weak. Even in this case, our calculations using a periodic defect model (similar to virtual crystal model for a disordered system) shows that electronic states near the band extrema can be significantly perturbed [see Figs.~\ref{fig;Bands3xSbnBi}(a) and \ref{fig;Bands3xSbnBi}(b)]. Fluctuations from this periodic impurity model will however affect the electronic states near the band extrema, sometimes localizing them near the defects if these fluctuations are strong enough. In this case a localized picture for the electronic states near the band extrema will be more meaningful. These states will not contribute much to transport. The states contributing to transport can be handled through a simple rigid band picture.

Experimentally, Pb$_{1-x}$Sb$_{x}$Te ($x$=0.25, 0.50, and 1.00\%) samples were reported to have $n$-type conductivity and degenerate doping.~\cite{jaworski} In a supercell description, these concentrations correspond roughly to (5$\times$5$\times$5), (4$\times$4$\times$4), and (3$\times$3$\times$3) supercells, respectively. For the largest supercell ($x$$\sim$0.25\%), we expect the impurity-induced bands to approach the host PbTe bands and a simple carrier doping of the host bands by electrons is a reasonable picture. However, for the (3$\times$3$\times$3) supercell corresponding to $x$$\sim$1.00\%, the band structure near the band gap region gets modified and the simple carrier doping picture starts to break down. Transport measurements carried out on these samples showed that increasing the Sb content ($x$) resulted in an increase in the carrier concentration and decrease in the magnitude of thermopower.~\cite{jaworski} This can be understood in terms of the calculated band structures presented above, where increasing the impurity (Sb) concentration results in a larger band splitting and smaller band gap. The electrical conductivity was also reported to increase with increasing $x$ from 0.25 to 0.50\%, which is consistent with the increase in the carrier concentration. It, however, decreases in going from $x$=0.50 to 1.00\%; this was ascribed to scattering.~\cite{jaworski}

\subsubsection{Simultaneous doping with monovalent and trivalent impurities}

Figures~\ref{fig;BandsPbTe}(e)$-$\ref{fig;BandsPbTe}(h) show the band structures of PbTe when simultaneously doped with Ag and Sb (or Bi). In our formation energy calculations (Sec.~III) we find that indeed it is energetically favorable to dope PbTe simultaneously with monovalent (Ag, Na, and K) and trivalent (Sb and Bi) impurities. In these cases, we also identify impurity-induced bands. We observe bands which are split off from the valence-band top and the conduction-band bottom. The lowest conduction bands which were pushed down in the presence of the trivalent atoms are found to be pushed up in energy (with respect to the VBM) and the band gap opens up at the $\Gamma$ point (in the sc BZ) for certain pair configurations. The upward shift of the trivalent impurity-induced band is larger when the separation between the impurities is smaller, which can be seen more clearly in Figs.~\ref{fig;BandsPbTe}(f) and \ref{fig;BandsPbTe}(h) for the case of (Ag,Bi).

\begin{figure}
\hspace*{-0.15in}
\includegraphics[width=3.2in,angle=0]{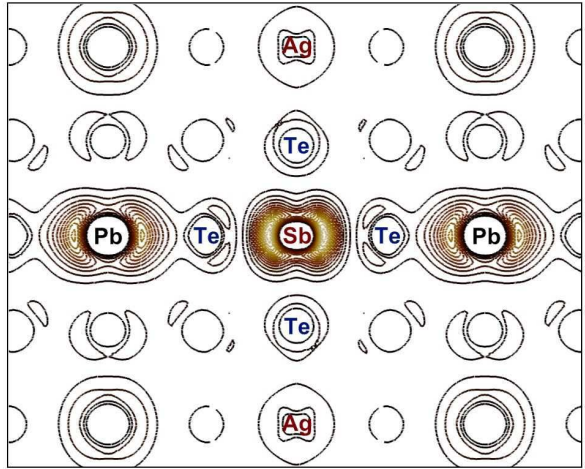} 
\caption{(Color online) Band-decomposed charge density associated with the Sb-induced band in (Ag,Sb)-doped PbTe. This partial density was calculated from a preconverged wavefunction file for the specified band. Contributions from all 18 $\mathbf{k}$-points in the irreducible BZ were included. The Sb-induced band is predominantly Sb $p$ and Pb $p$ states with some contribution from the $p$ states of the Te atoms which are the nearest-neighbor atoms of Sb. The hybridized $p$ states coming from the two Te atoms in the Ag-Te-Sb-Te chain and part of Sb are stabilized by Ag.} \label{fig;AgSb2IBSb}
\end{figure}

The introduction of Ag (in the presence of Sb or Bi) to PbTe, therefore, has an effect opposite to that of Sb(Bi). Since the tendency to drive the system towards metallicity when only Sb and Bi are present depends on the interaction between Sb(Bi) $p$ and Te $p$ (which are hybridized with the neighboring Pb $p$), the addition of Ag stabilizes the hybridized Te $p$ states and pushes the Sb(Bi)-induced band back towards the other conduction bands. An examination of the partial charge densities associated with the Ag-induced band [the highest valence band in Fig.~\ref{fig;BandsPbTe}(e)] and the Sb-induced band [the lowest conduction band in Fig.~\ref{fig;BandsPbTe}(e)] in PbTe simultaneously doped with Ag and Sb shows that this is indeed the case. The contour plot shown in Fig.~\ref{fig;AgSb2IBSb} clearly indicates that the hybridized $p$ states coming from the two Te atoms in the Ag-Te-Sb-Te chain are stabilized by Ag. This results in lowering the total energy of the system. The effect is strong for configurations with small pair distances and when the monovalent atom is in the direction of Sb(Bi) $p$-Te $p$ interaction, which is consistent with the energy landscapes for impurity pairs made of monovalent and trivalent atoms reported in Sec.~III.

Band gaps (at $\Gamma$) in the case of (Ag,Sb) are 94.1, 78.2, 81.5, 83.3 meV when the two atoms in the pair are at the first, second, third, and fifth n.n., respectively. For comparison, the calculated band gap for bulk PbTe is 105 meV. For (Ag,Bi), the gaps at $\Gamma$ are 43.4 and 0 meV for the second and fifth n.n.~distances, respectively, and negative for other distances. Since the splitting of the conduction band is caused by the spin-orbit part of the impurity potential, this splitting is larger and hence the band gap is smaller (or even negative) in the (Ag,Bi) case because SOI is much stronger for Bi. Also DFT-GGA underestimates the band gap of pure PbTe, it is most likely that the actual gaps in the presence of (Ag,Sb) and (Ag,Bi) pairs should be larger than those given here.

Experimentally, diffuse reflectance measurements carried out on Ag$_{1-x}$Pb$_{18}M$Te$_{20}$ ($M$=Bi and Sb) samples give an apparent gap of 0.25 eV for $M$=Bi and 0.28 eV for $M$=Sb;~\cite{LBST} both for $x$=0. The difference of $\sim$30 meV in the band gap is consistent with the above calculated values when the two atoms in a pair are at the second n.n.~distance; 43.4 meV for (Ag,Bi) and 78.2 meV for (Ag,Sb). If the conduction-band states are shifted rigidly upward in energy by $\sim$0.2 eV (to correct for DFT-GGA), then there appear to be a good agreement between experiment and our theoretical calculations.

In the current model ($M$$M'$Pb$_{30}$Te$_{32}$), the impurity concentration is $\sim$3\%, whereas in experiments, typically it is $\sim$5\%.~\cite{LAST,LBST} One, therefore, should expect that a monovalent atom (Ag) can easily find itself close enough to a trivalent atom (Sb or Bi) to have an impact on the latter. In addition, our first-principles study of the energetics of different pairs of impurities (see Sec.~III) also shows that an impurity pair made of a monovalent atom and a trivalent atom is most stable when the two atoms are the second n.n.~of one another. There is, however, still a finite chance for an impurity atom to be isolated or closer to another atom of the same kind, depending on the actual distribution of the impurity atoms in the samples and the ratio between the monovalent and trivalent atoms. This suggests that one can tune the band structure (hence transport properties) of the doped PbTe systems by tuning the Ag/Sb(Bi) ratio.

\subsubsection{Tuning the transport properties via the Ag/Sb(Bi) ratio}

Transport properties measurements carried out on $n$-type Ag$_{1-x}$Pb$_{18}$SbTe$_{20}$ ($x$=0, 0.14, and 0.30) by Han {\it et al.}~\cite{LBST} showed that the electrical conductivity increases and the absolute value of the thermopower decreases with decreasing Ag concentration (i.e., with increasing $x$) at a given temperature. The experimental data clearly indicate that the transport properties of the system depend on the Ag/Sb ratio. The magnitude of thermopower is largest and the electrical conductivity is smallest when Sb/Ag=$1$. We now discuss how these data can be understood in terms of the calculated band structures.

Let us look at the physics of the above situation using our band structure calculations. The stoichiometric compound ($x$=0) is charge-compensated and should be a semiconductor. The electron doping in this case comes most likely from intrinsic defects like Te vacancies. Concentration of such native defects is however small ($\sim$parts/million). Although Te vacancies do perturb the conduction-band bottom significantly,~\cite{hoangthesis} in this very dilute limit we do not expect a significant change in the band structure of PbTe. The concentrations of Ag and Sb are, on the other hand, sufficiently large such that changing the Ag/Sb ratio not only changes the electrons donated to the network but, more significantly, changes the band structure near the band gap. The observed behavior of the electrical conductivity and thermopower can be understood as follows: Increasing the Ag content increases the number of (Ag,Sb) pairs with short pair distances and reduces the number of isolated Sb impurities. This results in the widening of the band gap and reduction in the active carrier concentration, which leads to a decrease in the electrical conductivity and increase in the magnitude of thermopower.

The Ag$_{1-x}$Pb$_{18}$BiTe$_{20}$ system was also reported to produce similar behavior.~\cite{LBST} The dependence of electrical conductivity and thermopower on $x$ is, however, weaker in the Bi analog. The difference between (Ag,Sb) and (Ag,Bi) in PbTe-based thermoelectrics will be discussed in more detail in the next section. Studies of Zhu {\it et al.}~\cite{zhu} on quenched AgPb$_{18}$Sb$_{1-y}$Te$_{20}$ ($y$=0.0, 0.1, 0.3, and 0.5) samples also showed that increasing the Sb content (i.e., with decreasing $y$) results in reduced electrical conductivity and increase in thermopower. Again, the magnitude of thermopower is largest and the electrical conductivity is smallest when Sb/Ag=$1$. In this case, the increase of the Sb/Ag ratio (up to 1) increases the probability of having (Ag,Sb) impurity pairs with the Ag and Sb atoms being close to one another and reduces the number of isolated Sb impurities.

\subsubsection{Why the Bi analog of LAST-$m$ is an inferior thermoelectric?}

\begin{figure}
\includegraphics[width=3.38in,angle=0]{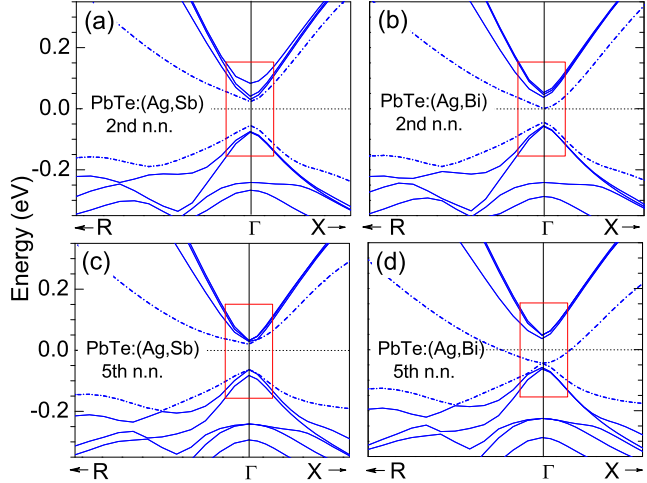} 
\vspace{-0.24in} \caption{(Color online) Band structures of PbTe doped with (Ag,Sb) and (Ag,Bi) impurity pairs showing the differences in the band gap and in the arrangement of the energy bands near the $\Gamma$ point. The impurity-induced bands are represented by the dash-dotted curves. The Fermi level (0 eV) is set to the highest occupied states or in the band gap.} \label{fig;SbvBi}
\end{figure}

In order to understand the difference between the two systems, (Ag,Sb)- and (Ag,Bi)-doped PbTe, we show in Figs.~\ref{fig;SbvBi}(a)$-$\ref{fig;SbvBi}(d) the blow ups of the band structures near the band gap at the $\Gamma$ point. Since LAST-$m$ and its Bi analog are $n$-type thermoelectrics (obtained by suitable adjustments of Ag concentration and the presence of unknown intrinsic defects, probably anion vacancies),~\cite{LAST,LBST} we focus on the region near the conduction-band bottom.

In addition to the difference in the band gap between these two systems, there are other differences in the ordering of the energy bands and their multiplicity. For (Ag,Sb) at the second n.n.~distance, there is a group of three bands (each band is a doublet when spin is included) which are close in energy at the $\Gamma$ point, see Fig.~\ref{fig;SbvBi}(a); $\sim$42 meV above the highest band in this group is another band (which is also a doublet). The splitting of the bands is much smaller for (Ag,Sb) at the fifth n.n.~distance, see Fig.~\ref{fig;SbvBi}(c). The arrangement of the bands in the (Ag,Bi) case is, however, in the reverse order; the doublet is below the sextet by $\sim$37 meV and the band splittings are larger, see Figs.~\ref{fig;SbvBi}(b) and \ref{fig;SbvBi}(d). This means that, for the same carrier (electron) concentration, the chemical potential in the (Ag,Sb) case is lower (closer to the CBM) than in the (Ag,Bi) case. This and the larger band gap for (Ag,Sb) which helps reduce the contribution from the minority carriers (holes) will result in a larger thermopower (magnitude) for (Ag,Sb)-doped PbTe.

Experimentally, transport properties measurements of Ag$_{1-x}$Pb$_{18}M$Te$_{20}$ ($M$=Bi and Sb) showed that thermopower decreases dramatically when Sb is replaced by Bi; from $-$100 $\mu$V/K at 300 K and $-$250 $\mu$V/K at 700 K for $M$=Sb to $-$40 $\mu$V/K at 300 K and $-$160 $\mu$V/K at 600 K for $M$=Bi.~\cite{LBST} The lattice thermal conductivity of the Bi analog is, however, larger than that of Sb because of the smaller mass fluctuation in this system (Bi and Pb are comparable, whereas Sb is much lighter than Pb). This higher thermal conductivity, coupled with the lower values of the thermopower, results in much lower $ZT$ in Ag$_{1-x}$Pb$_{18}$BiTe$_{20}$, $ZT$=0.44 at 665 K ($x$=0.3), compared to $ZT$$\sim$1.0 at 650 K for its Sb analog.~\cite{LBST}

Finally, looking at the calculated band structures presented in Figs.~\ref{fig;BandsPbTe}(e)$-$\ref{fig;BandsPbTe}(h), we expect that $p$-type LAST-$m$, if it can be successfully synthesized, will give a large thermopower. As far as the band structure is concerned, the thermopower can be even larger for the $p$-type system than the $n$-type one because the energy bands near the VBM (which are predominantly the hybridized states of Ag $d$ and Te $p$) are flatter than those near the CBM. Likewise, the $p$-type Bi analog may give larger thermopower than the $n$-type one.

\subsubsection{Why Na and K are good substitutes for Ag?}

\begin{figure}
\hspace*{-0.1in}
\includegraphics[width=3.4in,angle=0]{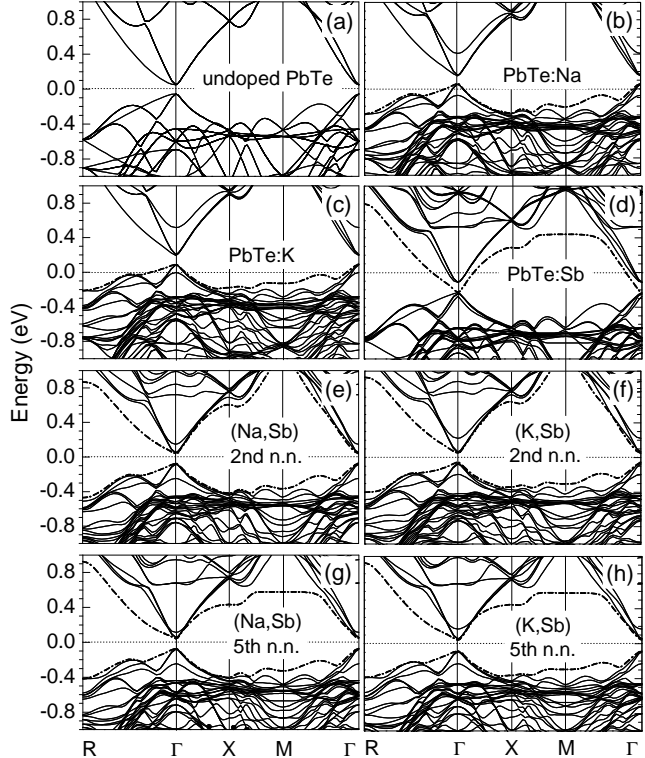} 
\vspace{-0.48in}
\caption{Band structures of undoped PbTe and PbTe doped with Na, K, Sb, (Na,Sb), and (K,Sb). The impurity-induced bands are represented by the dash-dotted curves. The results were obtained in calculations using the (2$\times$2$\times$2) supercell with SOI included. The two impurity atoms in a pair are either the second (2nd n.n.) or fifth (5th n.n.) nearest neighbors of one another. The Fermi level (0 eV) is set to the highest occupied states or in the band gap.}
\label{fig;BandsSoPo}
\end{figure}

The substitution of Ag in LAST-$m$ by Na or K have shown that these Ag-free thermoelectrics are also very promising.~\cite{SALT,PALT} The $p$-type Na$_{1-x}$Pb$_{m}$Sb$_{y}$Te$_{m+2}$ gives $ZT$$\sim$1.7 at 650 K for $m$=20,
$x$=0.05, and $y$=1,~\cite{SALT} whereas the $n$-type K$_{1-x}$Pb$_{m+\delta}$Sb$_{1+\gamma}$Te$_{m+2}$ gives $ZT$$\sim$1.6 at 750 K for $m$=20, $x$=0.05, $\delta$=0, and $\gamma$=0.2.~\cite{PALT} These systems are good thermoelectrics because of their large thermopower and low thermal conductivity. The latter is believed to be due to the nanostructuring observed in the systems,~\cite{SALT,PALT} which is also consistent with the our energetic studies presented in Sec.~III. In order to understand the large thermopower values observed experimentally, we have analyzed the electronic structures of PbTe doped with Na, K, (Na,Sb), and (K,Sb). The calculated band structures are presented in Figs.~\ref{fig;BandsSoPo}(a)$-$\ref{fig;BandsSoPo}(h).

Let us first examine the band structure for a single Na or K substituting for a Pb in a (2$\times$2$\times$2) cubic supercell model. Although thought to be ideal acceptors,~\cite{ahmadPRB} Na and K indeed produce significant changes in the band structure of PbTe, with band splittings near the valence-band top and conduction-band bottom, see Figs.~\ref{fig;BandsSoPo}(b) and \ref{fig;BandsSoPo}(c). The VBM (CBM) of PbTe, which originally has eightfold degeneracy, now splits into a group of three nearly degenerate bands and a stand alone band (a doublet) at the
$\Gamma$ point. The band nominally associated with Na (or K) [the dash-dotted bands in Figs.~\ref{fig;BandsSoPo}(b) and \ref{fig;BandsSoPo}(c)], in fact, does not have any Na (K) character since the Na (K) $s$ level is high up in the conduction band. It is, however, formed primarily out of $p$-orbitals associated with Te atoms which are the n.n.~of the Na (K) atom. This band will still be called the ``impurity-induced band'' since it is induced by Na(K) and split off from the top-most valence band and the rest of the PbTe valence bands.

The band structure for Sb-doped PbTe has already been discussed but is given in Fig.~\ref{fig;BandsSoPo}(d) for comparison. Simultaneous doping with Sb and Na (or K) helps push the Sb-induced bands upwards in energy, resulting in a band gap at $\Gamma$. Na (K) acts just like Ag in this regard, i.e, stabilizing the hybridized Te $p$ states resulting from the interaction between Sb $p$ and Te $p$. The Sb-induced band is, however, pushed to higher energy in the case of Na (and K), compared to Ag; and the new band gap can be even be larger than that of the undoped
PbTe; 114 meV and 112 meV for the (Na,Sb) and 100 meV and 122 meV for the (K,Sb) pair at the second and the fifth n.n.~distances, respectively, as compared to 105 meV for the undoped PbTe. This is qualitatively consistent with experiment where K$_{1-x}$Pb$_{m+\delta}$Sb$_{1+\gamma}$Te$_{m+2}$ ($m$=19, 20, and 21) was found to have ($\sim$0.03 eV) larger band gap than pure PbTe.~\cite{PALT} This suggests that Na and K, which are more ionic than Ag, compensate almost completely the charge perturbation created by Sb. In addition, K and Sb atoms may come close and locally form KSbTe$_{2}$ (whose band gap is $\sim$0.55 eV),\cite{PALT} which may explain the increased band gap of K$_{1-x}$Pb$_{m+\delta}$Sb$_{1+\gamma}$Te$_{m+2}$.

\begin{figure}
\includegraphics[width=3.35in,angle=0]{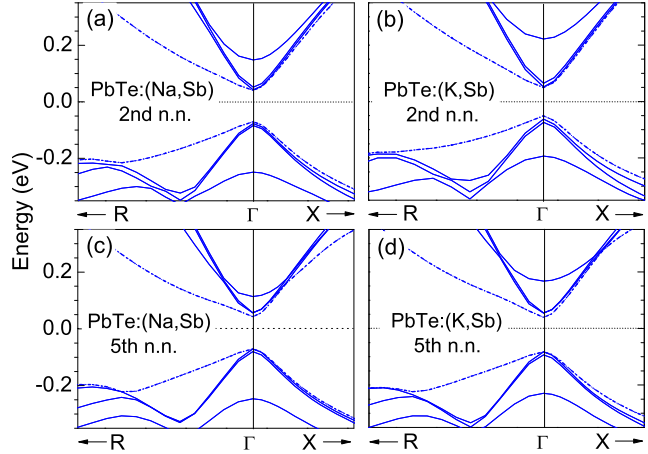} 
\vspace{-0.10in}
\caption{(Color online) Band structures of PbTe doped with (Na,Sb) and (K,Bi) impurity pairs showing the arrangement of the energy bands near the $\Gamma$ point. The impurity-induced bands are represented by the dash-dotted curves. The Fermi level (0 eV) is in the band gap.}
\label{fig;SoPoBlowups}
\end{figure}

There are other differences in the band structures of PbTe doped with Na, K, and Ag. The splitting between the threefold nearly degenerate band and the nondegenerate band near the conduction-band bottom at the $\Gamma$ point increases in going from Ag [Figs.~\ref{fig;SbvBi}(a) and \ref{fig;SbvBi}(c)] to Na [Figs.~\ref{fig;SoPoBlowups}(a) and \ref{fig;SoPoBlowups}(c)] to K [Figs.~\ref{fig;SoPoBlowups}(b) and \ref{fig;SoPoBlowups}(d)]; see also Table~\ref{tab;splitting}. The trend in the splitting near the conduction-band bottom reflects the positions of the $s$ level of Ag, Na, and K given by the Harrison's table; which are respectively at $-$5.99 eV (Ag), $-$4.96 eV (Na), and $-4.01$ eV (K).~\cite{harrisonTable} The trend in the splitting near the valence-band top is probably due the difference in the dopants (Ag, Na, and K) and the host (Pb) potentials. Since the difference in the splitting between Ag, Na, and K is small near the valence-band top and large near the conduction-band bottom, one expects that the difference in the thermopower (at a given carrier concentration) is small for $p$-type systems but larger for the $n$-type ones.

\begin{table}
\caption{The splitting (in meV) between the group of three nearly degenerate bands and the single band at the conduction-band minimum (CBM) and valence-band maximum (VBM) (at the $\Gamma$ point in the simple cubic BZ) of PbTe simultaneously doped with Sb and either Ag, Na, or K. The two atoms in a pair are either the second or fifth n.n.~of one another. The results were obtained in calculations using the (2$\times$2$\times$2) supercell with SOI included.} \label{tab;splitting}
\begin{center}
\begin{tabular}{ccccccc}
\hline \hline
&\multicolumn{2}{c}{(Ag,Sb)}&\multicolumn{2}{c}{(Na,Sb)}&\multicolumn{2}{c}{(K,Sb)}\\ \hline
&2nd n.n.&5th n.n. &2nd n.n.&5th n.n. &2nd n.n.&5th n.n. \\ \hline
CBM&42&0&96&58&157&112\\ 
VBM&167&160&165&166&132&136\\
\hline \hline
\end{tabular}
\end{center}
\end{table}

Experimentally, Na$_{0.8}$Pb$_{20}$Sb$_{y}$Te$_{22}$ compositions with $y$=0.4, 0.6, and 0.8 was found to have the ($p$-type) electrical conductivity decrease and thermopower increase as one increases the Sb content ($y$).~\cite{SALT} The thermopower is largest and the electrical conductivity is smallest when Sb/Na=$1$. This is similar to what has been observed in the case of Ag (see Sec.~IV.A.3). In the Na analog for $y$=0.8, one can assume that Pb vacancies are responsible for the observed $p$-type conductivity. Note that Pb vacancies in PbTe act like the monovalent impurities, i.e., introducing a similar ``impurity-induced band'' near the valence-band top and turn PbTe into a $p$-type system.~\cite{hoangthesis} For $y$$<$0.8, the $p$-type conductivity in the Na analog could be due to the excess Na (over Sb). The behavior of the electrical conductivity and thermopower can then be understood in terms of the calculated band structures where the increase of the Sb/Na ratio (up to 1) increases the probability of having (Na,Sb) pairs with short pair distances and reduces the isolated Sb impurities. This is similar to what has been discussed for Ag.

\subsection{GeTe}

One of the best materials today for power generation is (AgSbTe$_{2}$)$_{1-x}$(GeTe)$_{x}$, known as TAGS (stands for Te-Ag-Ge-Sb), which is a $p$-type thermoelectric and has $ZT$ as high as 1.36 at 700 K (for $x$=0.85).~\cite{wood,salvador} This system is thought to be a solid-solution of AgSbTe$_{2}$ and GeTe. Its Bi analog, (AgBiTe$_{2}$)$_{1-x}$(GeTe)$_{x}$, has also been found to be a good ($p$-type) thermoelectric material with $ZT$=1.32 at 700 K (for $x$=0.97).~\cite{plachkova} To understand the role Ag and Sb (Bi) play in these systems, band structure calculations were carried out where Ag and Sb (Bi) were treated as impurities in GeTe.

\begin{figure}
\hspace*{-0.15in}
\includegraphics[width=3.5in,angle=0]{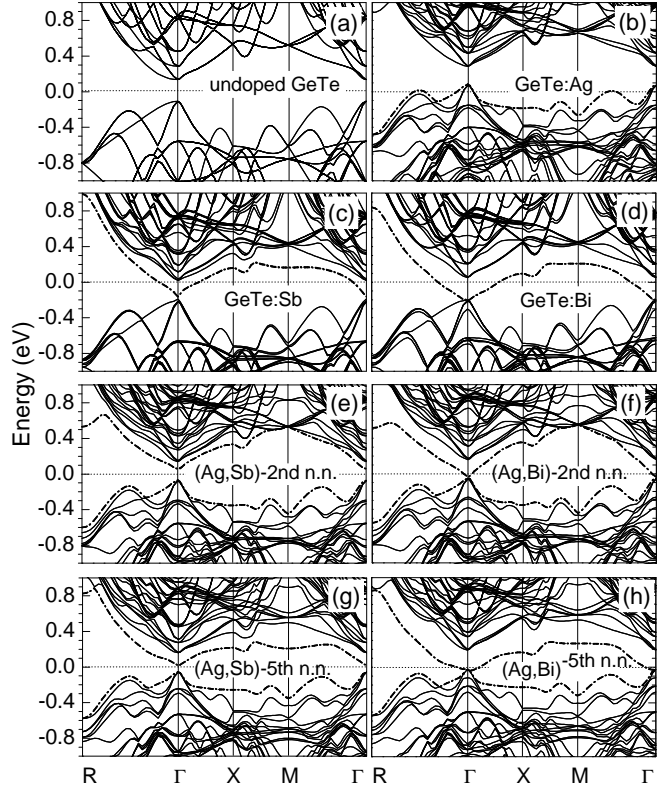} 
\vspace{-0.55in} \caption{Band structures of undoped GeTe and GeTe doped with Ag, Sb, Bi, (Ag,Sb), and (Ag,Bi). The impurity-induced bands are represented by the dash-dotted curves. The results were obtained in calculations using the (2$\times$2$\times$2) supercell with SOI included. The two impurity atoms in a pair are either the second (2nd n.n.) or fifth (5th n.n.) nearest neighbors of one another. The Fermi level (0 eV) is set to the highest occupied states or in the band gap.} \label{fig;BandsGeTe}
\end{figure}

Let us first look at the undoped system. Figure~\ref{fig;BandsGeTe}(a) shows the band structure of undoped fcc GeTe along different high symmetry directions of the sc BZ. This band structure, which was obtained using the (2$\times$2$\times$2) supercell (64 atoms/cell) with the optimized lattice constant of GeTe, looks very much like that of PbTe as far as the highest valence bands and the lowest conduction bands are concerned. There are, however, major differences in the positions of the higher conduction bands.

The calculated band gap (at the $\Gamma$ point) is 0.243 eV, see Fig.~\ref{fig;BandsGeTe}(a); $\sim$20\% larger than the reported value (of $\sim$0.20 eV) in experiment.~\cite{chang} Without SOI, our calculations give GeTe a band gap of 0.388 eV. The reduction of the band gap (by 37.4\%) due to SOI is caused by a small upward shift (by $\sim$0.050 eV) in the energy of the highest valence band and the spin-orbit-induced splitting of the lowest conduction band where one of the split bands is pushed down (by $\sim$0.096 eV) in energy. We note that the lattice distortion is not present in our current calculations. It is most likely that if we take into account the effects of lattice distortion the calculated band gap will be smaller than that of experiment, consistent with the known limitations of DFT-GGA.

The introduction of Ag into GeTe (Ag substitutes for Ge) results in a $p$-type system with the Fermi level below the VBM. Ag-induced band is the one that splits off from the valence-band top (of GeTe), presented by the dash-dotted curve in Fig.~\ref{fig;BandsGeTe}(b). Sb and Bi, on the other hand, make GeTe $n$-type. Impurity-induced bands associated with Sb and Bi which are split off from the GeTe conduction-band bottom significantly reduce the band gap (in the case of Sb) or even close the gap (Bi), see Figs.~\ref{fig;BandsGeTe}(c) and \ref{fig;BandsGeTe}(d). These observations are similar to what was seen for Ag, Sb, and Bi in PbTe, except that band crossing is less severe for GeTe because it has a larger band gap.

Simultaneous doping of Ag and Sb (or Bi) also helps lift the impurity-induced band associated with Sb (or Bi) up in energy and increases (or even opens up) the gap, see Figs.~\ref{fig;BandsGeTe}(e)$-$\ref{fig;BandsGeTe}(h); again similar to what was seen in PbTe [Figs.~\ref{fig;BandsPbTe}(e)$-$\ref{fig;BandsPbTe}(h)]. The arrangement of the energy bands near the CBM of (Ag,Sb) doped GeTe is, however, in the reverse order compared to that in PbTe. The nondegenerate (disregarding spins) band is now at a lower energy than the nearly threefold degenerate band. This is because the nondegenerate band is spin-orbit-induced and the effects is much larger for Sb and Bi than for Ge.

Although (Ag,Sb)- and (Ag,Bi)-doped GeTe have the same band arrangements near the CBM, the splitting between the threefold degenerate band and the nondegenerate impurity-induced band is smaller in the (Ag,Sb)-doped system; 82.8 meV and 143.8 meV at the second and fifth n.n.~distances, respectively, compared to 194.9 meV and 219.4 meV in the (Ag,Bi)-doped system. For $n$-type thermoelectrics, contribution to the transport properties largely comes from the lowest conduction bands. The (Ag,Sb)-doped system is, therefore, expected to give higher thermopower compared to the (Ag,Bi)-doped one, as far as the band structures are concerned, because the former has a larger band gap.

The arrangement of the energy bands near the conduction-band bottom is expected not to affect the $p$-type system much. However, the larger band gap should also result in larger thermopower in the (Ag,Sb)-doped case, compared to the (Ag,Bi); again, due to reduced intrinsic contributions in the (Ag,Sb)-doped system. The system with Bi, on the other hand, has lower lattice thermal conductivity because of the larger mass difference between Bi and Ge. Taking into account these different factors, one might expect that the $p$-type TAGS system and its Bi analog, if made under the same synthesis conditions, have comparable $ZT$ values.

\subsection{SnTe}

Besides the PbTe- and GeTe-based systems, complex quaternary systems based on SnTe have also shown to be promising for thermoelectric applications. Androulakis {\it et al.}~\cite{TAST} have reported an unusual coexistence of large thermopower and degenerate doping in the nanostructured material Ag$_{1-x}$SnSb$_{1+x}$Te$_{3}$ ($x$=0.15). This system shows a positive thermopower of $\sim$160 $\mu$V/K at 600 K and an almost metallic carrier concentration of $\sim$5$\times$10$^{21}$ cm$^{-3}$. In order to fully understand the role Ag and Sb play in these and other SnTe-based systems, especially those at low impurity concentration (lightly doped) limit ($\sim$3\%), band structure calculations were also carried out where Ag and Sb were treated as defects in SnTe.

\begin{figure}
\hspace*{-0.10in}
\includegraphics[width=3.5in,angle=0]{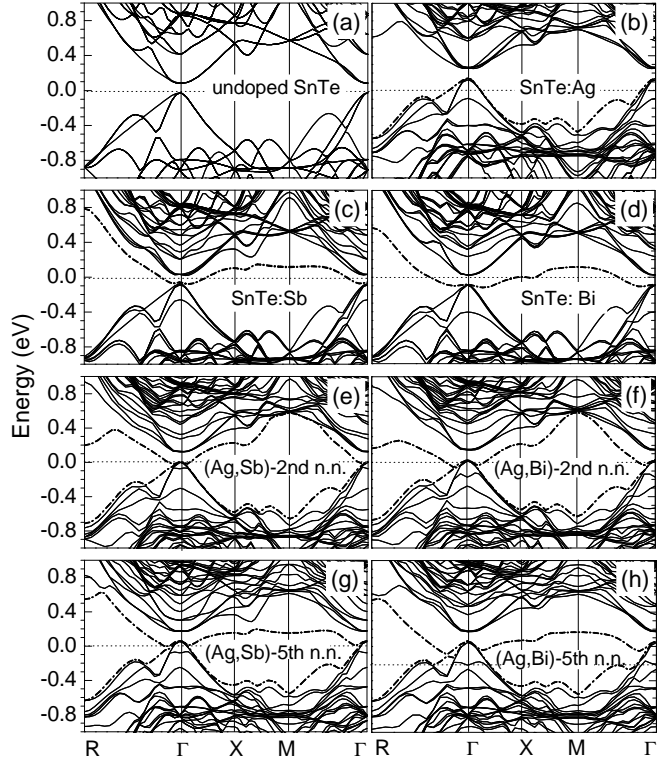} 
\vspace{-0.7in} \caption{Band structures of undoped SnTe and SnTe doped with Ag, Sb, (Ag,Sb), and (Ag,Bi). The impurity-induced bands are represented by the dash-dotted curves. The results were obtained in calculations using the (2$\times$2$\times$2) supercell with SOI included. The two impurity atoms in a pair are either the second (2nd n.n.) or fifth (5th n.n.) nearest neighbors of one another. The Fermi level (at 0 eV) is set to the highest occupied states or in the band gap.} \label{fig;BandsSnTe}
\end{figure}

It is well known that the electronic structure of SnTe is anomalous compared to that of PbTe.~\cite{cohen} The main differences between these two systems are: (i) SnTe has a direct gap which is slightly away from the $L$ point (along $L$-$W$) of the fcc BZ if SOI is included, but is a zero-gap semiconductor if SOI is excluded; whereas PbTe has a direct gap at the $L$ point even in the absence of SOI. (ii) The symmetry of the band-edge states at and near the $L$ point changes in going from PbTe to SnTe, resulting in the so-called ``band inversion'' phenomenon in SnTe.~\cite{cohen,tsang,dantas} Figure~\ref{fig;BandsSnTe}(a) shows the band structure of undoped SnTe along high symmetry directions of the sc BZ, obtained in calculations using the (2$\times$2$\times$2) supercell. In the sc BZ, the band gap of SnTe occurs near the $\Gamma$ point. The $\Gamma$ point is actually a saddle point, although the feature is not so evident from Fig.~\ref{fig;BandsSnTe}(a) where the band structure is plotted only along $\Gamma$-$R$ and $\Gamma$-$X$. The calculated band gap at $\Gamma$ is $\sim$0 eV (without SOI) and 0.105 eV (with SOI). Experimentally, SnTe was reported to have a band gap of 0.3 eV at 0 K.~\cite{esakiSnTe}

The band structure of SnTe doped with Ag is shown in Fig.~\ref{fig;BandsSnTe}(b). The impurity-induced band associated with Ag [the dashed-dotted curve in Fig.~\ref{fig;BandsSnTe}(b)] is predominantly Ag $d$ and Te $p$ states, splitting off from the valence-band top. The impurity-induced band associated with Sb [the dash-dotted curve in Fig.~\ref{fig;BandsSnTe}(c)], on the other hand, splits off from the conduction-band bottom and is predominantly Sb $p$ and Sn $p$ with some contribution from the hybridized $p$ states of the n.n.~Te atoms of Sb. This hybridized impurity-induced band comes down and closes the gap. These observations are similar to those in PbTe and GeTe containing Ag and Sb. There are, however, noticeable differences near the conduction-band bottom. The saddle-point feature of $\Gamma$ is more pronounced (compared to undoped SnTe) along the $\Gamma$-$R$ and $\Gamma$-$X$ directions. The separation between the maximum at $\Gamma$ and the VBM is $\sim$3 meV for SnTe doped with Ag.

The electronic structure in the case of simultaneous doping is also different from what has been observed in PbTe and GeTe. The introduction of Ag does lift the Sb- and Bi-induced bands up in energy, but not the part near the $\Gamma$ point; see Figs.~\ref{fig;BandsSnTe}(e)$-$\ref{fig;BandsSnTe}(h). It, actually, seems to pull the impurity-induced bands near the $\Gamma$ point further down in energy resulting in a larger overlap with the valence band. Based on the calculated band structures, one expects to see the (Ag,Sb)-doped SnTe systems to exhibit highly degenerate semiconducting or even metallic conduction. However, large thermopower for the $p$-type SnTe-based systems is also possible because of the high degeneracy of the bands near the VBM. This is consistent with the observation of large thermopower and high carrier concentration in experiment.~\cite{TAST} We note that the overlap between the impurity-induced bands and the valence band in these SnTe-based systems can be smaller or the systems may actually have small but finite gaps if one makes corrections to DFT-GGA calculations.

\section{Summary}

In summary, the energetics of impurity clustering and the impurity-induced bands associated with various monovalent (Na, K, and Ag) and trivalent (Sb and Bi) impurities in PbTe, SnTe, and GeTe were studied using first-principles calculations and supercell models. Our results showed that the impurity pairs formed out of monovalent and trivalent atoms in PbTe, SnTe, and GeTe have the lowest energy when they are the second nearest neighbors of one another. The Te atoms near an impurity pair (and some impurity atoms themselves) tend to go ``off-center,'' mainly because of the strong and directional interaction between the $p$ states of the trivalent impurity cations (Sb and Bi) and the divalent anion (Te). One should carry out experimental studies to see if there is indeed off-centering in the real samples and how it impacts the transport properties of these systems.

The electronic structures of PbTe, SnTe, and GeTe are strongly perturbed by the impurities (in the concentration range $>$1\%) and the degree of perturbation depends on the relative separation between the impurities. Band degeneracy at the VBM and CBM of the host materials is removed and there are band splittings in the presence of an impurity. We found that the impurity-induced bands associated with the monovalent impurities (Na, K, and Ag) split off from the valence-band top but overlap with the other valence bands near the VBM. All these bands, however, constitute the valence band and the systems are hole-doped. The impurity-induced bands associated with the trivalent impurities (Sb and Bi), on the other hand, split off from the conduction-band bottom with large shifts towards the valence band. These bands reduce or even close the band gap. This is due to the strong interaction between the $p$ states of the trivalent impurity cation (Sb and Bi) and the divalent anion (Te) which tends to introduce electronic states in the band gap region and drive the systems towards metallicity.

The simultaneous doping of monovalent and trivalent impurities, however, pushes the impurity-induced band associated with the trivalent impurity to higher energies and the band gap increases or opens up, except for the SnTe-based systems. One, therefore, can tune the monovalent/trivalent ratio to tune the band gap and the band structure near the band edges in PbTe and GeTe-based systems. SnTe behaves rather anomalously in the sense that the introduction of the monovalent impurity does not lift the trivalent impurity-induced band near the CBM up in energy but slightly pulls it further down towards the valence band. Based on the calculated band structures, we have been able to explain qualitatively the observed transport properties of the whole class of PbTe-, SnTe-, and GeTe-based bulk thermoelectrics.

In spite of the known limitations of DFT-GGA and supercell methods,~\cite{supercellmethod} our first-principles studies thus provide a clear physical picture of how different impurities affect the lattice geometry and electronic structure of the host materials. These results provide a qualitative understanding of the atomic and transport properties of complex multicomponent PbTe-, SnTe-, and GeTe-based thermoelectrics. The approach described here however is not limited to these systems but can be extended to other classes of materials. The ``band-gap problem'' can, in principle, be addressed by using more advanced methods which go beyond DFT-GGA.~\cite{sxLDA,HSE,aulber} These calculations for our systems are currently too demanding computationally; however, they are very desirable since having accurate electronic structures of the materials is an important step towards a quantitative understanding of their transport properties.

\begin{acknowledgments}

This work was partially supported by the U.S.~Office of Naval Research and made use of the computing facilities of Michigan State University High Performance Computing Center.

\end{acknowledgments}


\begin{references}
\bibitem{note} We note that, even within DFT-GGA, our calculations have already required an excessive amount of computer memory and processing time. A typical calculation (with spin-orbit interaction included) requires from several gigabytes of memory for the (2$\times$2$\times$2) supercell to several tens of gigabytes of memory for the (3$\times$3$\times$3) one. It takes about one week on a SGI Altix 3700 Bx2 machine to finish a self-consistent run that generates charge density, e.g., for Sb-doped PbTe using the (3$\times$3$\times$3) supercell.
\end{references}

\end{document}